\documentclass{emulateapj}

\def\hda{\mbox{HD~108317}}

\def\hdd{\mbox{HD~128279}}
\def\bd{\mbox{BD~$+$17~3248}}

\def\kmsec{\mbox{km~s$^{\rm -1}$}}
\def\logg{\mbox{log~{\it g}}}

\def\teff{\mbox{$T_{\rm eff}$}}
\def\vt{\mbox{$v_{\rm t}$}}
\def\rpro{\mbox{$r$-process}}
\def\spro{\mbox{$s$-process}}

\shorttitle{Detection of the Second $r$-process Peak}
\shortauthors{Roederer et al.}
\slugcomment{Accepted for publication in the Astrophysical Journal Letters}

\begin{document}

\title{
Detection of the Second $r$-process Peak Element Tellurium
in Metal-Poor Stars\footnotemark[1]$^{~ \rm ,}$\footnotemark[2] 
}

\footnotetext[1]{
Based on observations made with the NASA/ESA Hubble Space Telescope, 
obtained at the Space Telescope Science Institute, 
which is operated by the Association of Universities for Research in 
Astronomy, Inc., under NASA contract NAS~5-26555. 
These observations are associated with programs 
GO-8342 and
GO-12268.}

\footnotetext[2]{
Some of the data presented in this paper were obtained from the 
Multimission Archive at the Space Telescope Science Institute (MAST). 
STScI is operated by the Association of Universities for Research in 
Astronomy, Inc., under NASA contract NAS5-26555. 
These data are associated with programs
GO-7348, 
GO-7402, 
GO-8197, and
GO-9455.}

\author{
Ian U.\ Roederer,\altaffilmark{3}
James E.\ Lawler,\altaffilmark{4}
John J.\ Cowan,\altaffilmark{5}
Timothy C.\ Beers,\altaffilmark{6,}\altaffilmark{7,}\altaffilmark{8}
Anna Frebel,\altaffilmark{9}
Inese I.\ Ivans,\altaffilmark{10}
Hendrik Schatz,\altaffilmark{7,}\altaffilmark{8,}\altaffilmark{11}
Jennifer S.\ Sobeck,\altaffilmark{12}
Christopher Sneden\altaffilmark{13}
}

\altaffiltext{3}{Carnegie Observatories, 
Pasadena, CA 91101, USA
}
\altaffiltext{4}{Department of Physics, University of Wisconsin, 
Madison, WI 53706, USA
}
\altaffiltext{5}{Homer L.\ Dodge Department of Physics and Astronomy,
University of Oklahoma, Norman, OK 73019, USA
}
\altaffiltext{6}{National Optical Astronomy Observatory, 
Tucson, AZ 85719, USA
}
\altaffiltext{7}{Department of Physics \& Astronomy,
Michigan State University, E.\ Lansing, MI  48824, USA
}
\altaffiltext{8}{Joint Institute for Nuclear Astrophysics, 
Michigan State University, E.\ Lansing, MI  48824, USA
}
\altaffiltext{9}{Massachusetts Institute of Technology, 
Kavli Institute for Astrophysics and Space Research, 
Cambridge, MA 02139, USA
}
\altaffiltext{10}{Department of Physics and Astronomy, University of Utah,
Salt Lake City, UT 84112, USA
}
\altaffiltext{11}{National Superconducting Cyclotron Laboratory, 
Michigan State University, East Lansing, MI 48824, USA
}
\altaffiltext{12}{Department of Astronomy \& Astrophysics, 
University of Chicago, Chicago, IL 60637, USA
}
\altaffiltext{13}{Department of Astronomy, University of Texas at Austin,
Austin, TX 78712, USA
}


\addtocounter{footnote}{13}

\begin{abstract}

Using near-ultraviolet spectra obtained with the
Space Telescope Imaging Spectrograph onboard the 
\textit{Hubble Space Telescope},
we detect neutral tellurium 
in three metal-poor stars enriched
by products of $r$-process nucleosynthesis, 
BD~$+$17~3248, HD~108317, and HD~128279.
Tellurium (Te, $Z =$~52) is found at the 
second $r$-process peak ($A \approx$~130)
associated with the $N =$~82 neutron shell closure, and it
has not been detected previously in Galactic halo stars.
The derived tellurium abundances 
match the scaled Solar System $r$-process distribution
within the uncertainties,
confirming the predicted second peak $r$-process residuals.
These results suggest that tellurium is predominantly produced
in the main component of the $r$-process,
along with the rare earth elements.

\end{abstract}

\keywords{
nuclear reactions, nucleosynthesis, abundances ---
stars: abundances ---
stars: individual (BD~$+$17~3248, HD~108317, HD~128279) ---
stars: Population II
}

\section{Introduction}
\label{intro}

In the quest to understand the production of the
heaviest elements in the universe, 
several invaluable datasets for testing nucleosynthesis models 
are available,
including the isotopic and 
elemental abundance patterns found in
meteorites, the Solar photosphere, and metal-poor halo stars.
All elements heavier than the iron group
are produced (at least in part) by neutron-capture reactions.
These reactions occur on timescales that are slow or rapid, relative to
the average $\beta$-decay timescales of unstable nuclei along
the reaction chain, and thus they are known as the
\spro\ and \rpro, respectively.

The Solar System (S.S.) \rpro\ distribution
is calculated as the ``residual'' 
in the S.S.\ isotopic abundance distribution
after subtracting the predicted \spro\ distribution
from the total abundances.
This may be calculated analytically
(e.g., \citealt{seeger65,cameron82,kappeler89})
or via simulations that rely on stellar evolution models
and nuclear reaction networks
(e.g., \citealt{arlandini99,bisterzo11}).
Nuclei with neutron numbers $N =$~50, 82, and 126 
have reduced neutron-capture cross sections, giving
rise to ``peaks'' in the abundance distributions.
Models that capture the essential physics of these processes
must be able to reproduce the nucleosynthetic yields, and the
neutron-capture peaks are highly sensitive probes of this physics.

The third \rpro\ peak elements
osmium through platinum (76~$\leq Z \leq$~78; $A \approx$~195) are
associated with the $N =$~126 shell closure.
Their strongest absorption lines in stellar spectra are found
in the near-ultraviolet (NUV) and cannot be observed from 
ground-based facilities.
Using spectra obtained with the
Goddard High Resolution Spectrograph onboard the 
\textit{Hubble Space Telescope} (\textit{HST}),
\citet{cowan96} and \citet{sneden98} reported the first
detections of these elements in metal-poor stars 
enriched by the \rpro.
The osmium and platinum abundances were 
consistent with the scaled S.S.\ \rpro\ distribution.
This result established the similarity
of the predicted S.S.\ \rpro\ distribution and the 
\rpro\ distribution in halo stars that
formed many Gyr before the condensation of the Solar nebula.
Subsequent detections of osmium and platinum in additional halo stars
using the Space Telescope Imaging Spectrograph (STIS) onboard \textit{HST}
by \citet{cowan02,cowan05}, \citet{sneden03}, 
\citet{roederer10b}, and \citet{barbuy11}
have reaffirmed this similarity.

\begin{deluxetable*}{cccccccc}
\tablecaption{Atomic Transition Probabilities for Te~\textsc{i}
\label{tetab}}
\tablewidth{0pt}
\tabletypesize{\scriptsize}
\tablehead{
\colhead{Wavenumber} &
\colhead{$\lambda_{\rm air}$} &
\colhead{E$_{\rm upper}$} &
\colhead{$J_{\rm upper}$} &
\colhead{E$_{\rm lower}$} &
\colhead{$J_{\rm lower}$} &
\colhead{Transition Probability} &
\colhead{$\log(gf)$} \\
\colhead{(cm$^{-1}$)} &
\colhead{(\AA)} &
\colhead{(cm$^{-1}$)} &
\colhead{} &
\colhead{(cm$^{-1}$)} &
\colhead{} &
\colhead{(10$^{6}$ s$^{-1}$)} &
\colhead{} }
\startdata
 46652.738 & 2142.8218 & 46652.738 & 1 & 0.000     & 2 & 231(37)    & $-$0.32 \\
 41946.238 & 2383.2772 & 46652.738 & 1 & 4706.500  & 0 & 30(5)      & $-$1.11 \\
 41902.026 & 2385.7920 & 46652.738 & 1 & 4750.712  & 1 & 60(10)     & $-$0.81 \\
 36094.861 & 2769.6596 & 46652.738 & 1 & 10557.877 & 2 & 0.90(0.15) & $-$2.51 \\
\enddata
\end{deluxetable*}

Absorption lines of the second \rpro\ peak elements
tellurium, iodine, and xenon (52~$\leq Z \leq$~54; $A \approx$~130),
associated with the $N =$~82 shell closure,
are not observable from the ground in halo stars.
Transitions from the ground state of
iodine and xenon all occur at wavelengths shorter than 2000\,\AA\
(e.g., \citealt{morton00}). 
Tellurium has not been detected in the Solar photosphere, and its
S.S.\ abundance is known only from studies of meteorites
(e.g., \citealt{anders82}). 
Predictions suggest that $\approx$~17--20\% of the tellurium in the S.S.\ 
originated in the \spro, while the remainder was presumably produced by 
the \rpro\
(e.g., \citealt{sneden08,bisterzo11}).
Contributions from charged-particle reactions 
(i.e., proton captures) to tellurium
isotopes are expected to be minimal
(e.g., \citealt{kappeler89}).
We are aware of three previous detections of tellurium 
beyond the S.S.: 
a single, blended line in
the F-type main sequence star Procyon \citep{yushchenko96};
the \spro\ rich planetary nebula \mbox{IC~418} \citep{sharpee07}; and
the chemically peculiar star \mbox{HD~65949} \citep{cowley10},
which is not ideal for 
detailed exploration of the $s$- or \rpro\ patterns
due to the chemical stratification that occurs in its warm atmosphere.

The tellurium abundance in metal-poor halo stars
offers an independent check on the predicted 
\rpro\ abundance of tellurium in the S.S.
It also constrains the conditions of the \rpro\ operating 
early in the Galaxy.
In this Letter we report the detection of an absorption line of
Te~\textsc{i} at 2385\,\AA\ in the atmospheres of
three \rpro\ enriched stars observed with STIS.
We combine the abundances derived from this line with
abundances of other heavy elements
derived from both space- and ground-based data to form a 
more complete \rpro\ abundance distribution.

\section{Observations}
\label{observations}

In Program GO-12268, we have used STIS \citep{kimble98,woodgate98}
to obtain new spectra of 4~metal-poor stars with modest or sub-solar 
levels of \rpro\ enrichment.
The spectra of two of these stars, \hda\ and \hdd, are
useful for abundance work at 2385\,\AA.
These data were taken using the E230M echelle grating and the
NUV Multianode Microchannel Array (MAMA)
detector, yielding a spectral resolution
(R~$\equiv \lambda/\Delta\lambda$) $\sim$~30,000 and
wavelength coverage from 2280--3115\,\AA.
Standard reduction and calibration procedures were used;
see \citet{roederer12} for details.
We also examine previous STIS observations obtained in Program GO-8342
\citep{cowan02} of the metal-poor \rpro\ rich standard star \bd.
The signal-to-noise ratios in the continuum near 2385\,\AA\
are $\sim$~100/1 for \hda\ and \hdd\ and slightly less in \bd.

\section{The Te~I 2385\,\AA\ Absorption Line}
\label{teline}

\hda\ and \hdd\ have nearly identical stellar parameters and 
metallicities (Section~\ref{analysis}), 
but the heavy elements in
\hda\ are enhanced by a factor of $\sim$~2--3 relative to \hdd.
In these two stars,
most differences in the NUV spectral region can be traced to
absorption by elements heavier than the iron group.
One such difference is observed at 2385.8\,\AA, where
the absorption in \hda\ is stronger than in \hdd.
We have searched 
the National Institute of Standards and Technology (NIST) 
Atomic Spectra Database \citep{ralchenko11}, 
the \citet{kurucz95} line lists,
and the \citet{morton00}
compilation of low excitation transitions of heavy elements
to identify the species responsible for this absorption.
These sources indicate that
the only probable transition is 
Te~\textsc{i} 2385.79\,\AA.
The first ionization potential of tellurium is high (9.01~eV),
so a substantial amount of neutral tellurium ($\sim$~50\%)
is present.

We calculate transition probabilities of four Te~\textsc{i} transitions
sharing the same upper level as the 2385.79\,\AA\ transition.
These values are listed in Table~\ref{tetab}.
We adopt energy levels from \citet{morillon75}, 
branching ratios from \citet{ubelis83}, and the laser-induced
fluorescence lifetime measurement of the upper level 
from \citet{bengtsson92}.
The $\log(gf)$ values have an uncertainty of about 16\%, where the 
largest source of uncertainty arises from the lifetime measurement.

Other relatively strong transitions of Te~\textsc{i} should
lie at 2142.82\,\AA\ and 2383.28\,\AA.
Detecting either of these features would strongly confirm our
Te~\textsc{i} detection based on the 2385.79\,\AA\ transition.
Unfortunately, the 2383.28\,\AA\ transition is buried within
an absorption feature (EW~$>$~0.5\,\AA) 
dominated by the Fe~\textsc{ii} line at 2383.25\,\AA, and it
is not useful for this purpose.
The 2142.82\,\AA\ transition is not covered in our data.
In the STIS archives we find high-resolution
spectra of two metal-poor stars,
\mbox{HD~94028} and \mbox{HD~140283},
that cover both the 2142\,\AA\ and 2385\,\AA\ lines.
Both stars are warmer than those in our sample
(e.g., \citealt{peterson11}),
so the fraction of tellurium present in the neutral state in these
stars is lower. 
No absorption is detected at 2385.79\,\AA\ in \mbox{HD~140283}, but
a very weak feature is detected in \mbox{HD~94028}.
In \mbox{HD~94028}, the Te~\textsc{i} feature predicted at 2142.82\,\AA\
is blended with an absorption feature at 2142.83\,\AA\
that may be due to Fe~\textsc{i}.
Our syntheses suggest that some of the absorption may be
due to Te~\textsc{i}, but it is impossible to derive a 
reliable abundance from it.
We are unable to 
substantiate our detection with additional Te~\textsc{i} lines,
so we proceed with caution.

\section{Abundance Analysis}
\label{analysis}

\begin{figure}
\begin{center}
\includegraphics[angle=0,width=3.5in]{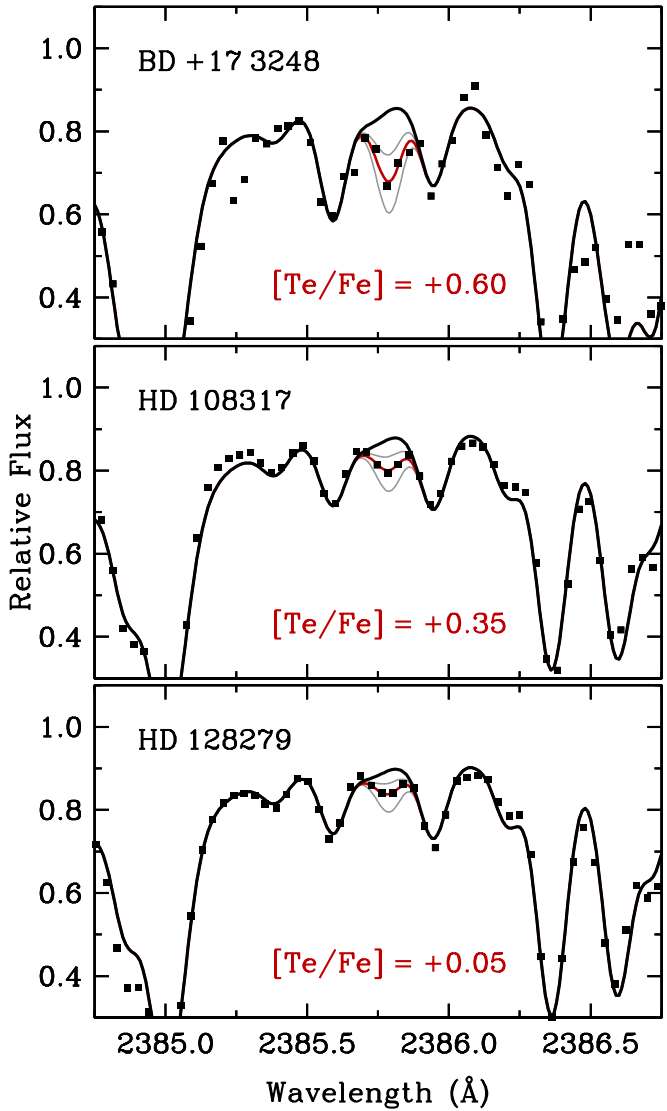}
\end{center}
\caption{
\label{specplot}
Comparison of observed and synthetic spectra around the Te~\textsc{i}
2385\,\AA\ line. 
The bold red line represents the best-fit abundance,
the thin gray lines represent variations in this abundance by $\pm$~0.30~dex,
and the bold black line represents a synthesis with no tellurium.
 }
\end{figure}

\begin{figure}
\begin{center}
\includegraphics[angle=0,width=3.5in]{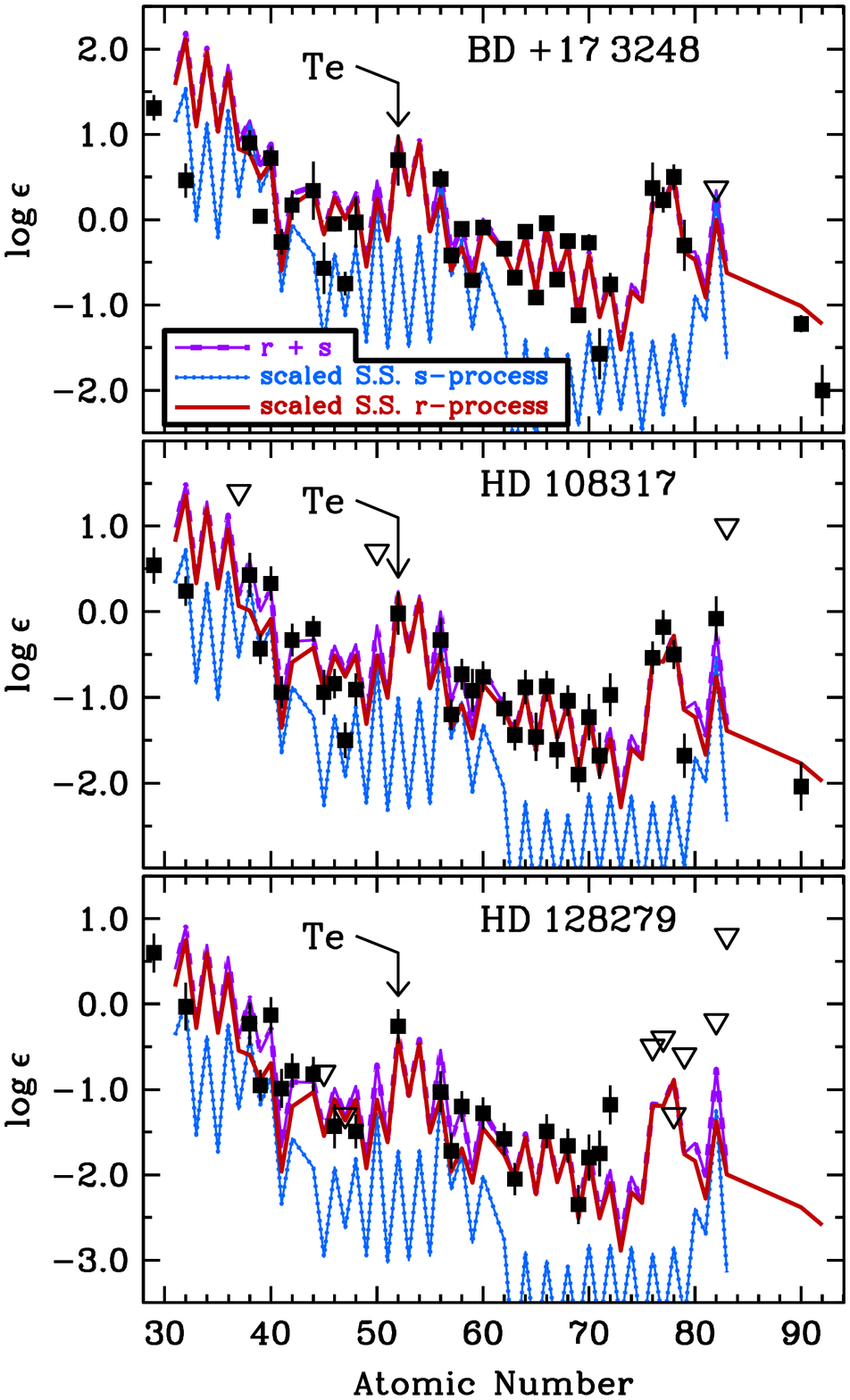}
\end{center}
\scriptsize
\caption{
\label{logepsplot}
Logarithmic abundances in \bd, \hda, and \hdd.
Filled squares indicate detections, and open downward facing triangles
indicate upper limits.
Two curves (blue and red) represent the scaled S.S.\ $s$-
and \rpro\ abundances from \citet{sneden08};
the ``strong'' \spro\ component contributions to
lead and bismuth are taken from \citet{bisterzo11}.
The third curve (purple) represents a combined $r+s$
distribution.
The \rpro\ and $r+s$ distributions are 
normalized to the europium (Eu, $Z =$~63)
abundance of each star, and the
\spro\ distributions are normalized to barium (Ba, $Z =$~56).
The tellurium abundances are derived in the present study.
References for the other abundances are given in the text.
}
\end{figure}

We perform a standard abundance analysis, using synthesis
techniques, to derive the tellurium abundance; see
\citet{roederer12} for details.
We use models interpolated from the 
$\alpha$-enhanced ATLAS9 grid \citep{castelli04}, and we
perform the analysis using the 
latest version of the analysis code MOOG \citep{sneden73}
under the assumption that local thermodynamic equilibrium (LTE)
governs the the excitation and ionization.
Our model parameters for \bd, \hda, and \hdd\ are, respectively,
\teff/\logg/[M/H]/\vt~$=$
5200~K/1.80/$-$2.08/1.90~\kmsec\ \citep{cowan02},
5100~K/2.67/$-$2.37/1.50~\kmsec, and
5080~K/2.57/$-$2.49/1.60~\kmsec\ \citep{roederer12}.

In the spectral region surrounding the Te~\textsc{i} 2385\,\AA\ line
($\pm$~10\,\AA),
our syntheses predict absorption from other species at 
approximately 95\% of
the wavelengths where a line is observed.
This agreement is reassuring.
Often the transition probabilities must be
adjusted to reproduce the observed spectrum, but
this is a well-known problem in the UV (e.g., \citealt{leckrone99}).

Our syntheses suggest 
that the Te~\textsc{i} line at 2385.79\,\AA\ is in a relatively clean
spectral window, as shown in Figure~\ref{specplot}.
Strong Fe~\textsc{ii} lines at 2383.04\,\AA\ and 2388.63\,\AA\ 
depress the continuum by 10--20\%,
which introduces additional 
uncertainty into the derived tellurium abundance.
We can fit
weaker Fe~\textsc{i} lines at 2385.59\,\AA\ and 2385.94\,\AA\
that mildly affect the observed Te~\textsc{i} line profile.
Our syntheses suggest that a Cr~\textsc{i} transition at 2385.72\,\AA\
contributes a small amount of absorption to the observed line profile.
NIST quotes an uncertainty of 40\% for this $\log(gf)$ value,
and we produce an acceptable fit to the overall line profile
by varying its strength accordingly.
We estimate errors by accounting for
uncertainties in the model atmosphere parameters, $\log(gf)$ values,
and the fit to the observed line profile
\citep{cowan05,roederer12}.
Zero-point uncertainties may exist
in abundance ratios of elements derived from species of different
ionization states.
We caution that such ratios
(e.g., [Te/Zr], [Te/Ba], [Te/Eu]) are less secure than
ratios derived from species in the same ionization state
(e.g., [Te/Pd], [Te/Pt], [Ba/Eu]).

We derive tellurium abundances ($\log\epsilon$)
from the 2385\,\AA\ line in
\bd, \hda, and \hdd\ of
$+$0.70~$\pm$~0.30,
$-$0.02~$\pm$~0.25, and 
$-$0.26~$\pm$~0.25, respectively.
Referencing these abundances to the iron abundance derived from
Fe~\textsc{i} in each star 
([Fe/H]~$= -$2.08, $-$2.55, $-$2.49, respectively)
and the S.S.\ meteoritic abundance of tellurium
($\log\epsilon = +$2.18; \citealt{asplund09}), we calculate
[Te/Fe]~$=+$0.60, $+$0.35, and $+$0.05, respectively.

\section{Discussion}
\label{results}

These stars are rich laboratories to study the nucleosynthesis
of heavy elements.
When the new tellurium abundances are
combined with previous abundance determinations in
\bd, \hda, and \hdd\ 
\citep{cowan02,cowan05,sneden09,roederer09,roederer10b,roederer12},
a total of 34, 34, and 22~elements heavier than zinc ($Z =$~30)
have been detected in the atmospheres of these stars.

Figure~\ref{logepsplot} illustrates the heavy element 
distributions in \bd, \hda, and \hdd.
The scaled S.S.\
$s$- and \rpro\ abundance patterns are shown for comparison
\citep{sneden08,bisterzo11}.
The \rpro\ residuals 
implicitly include contributions from all heavy element
nucleosynthesis mechanisms other than the \spro; e.g., a
light element primary process (LEPP; \citealt{travaglio04}).
The stellar abundance patterns---including the
new tellurium abundances---conform more closely to the
scaled S.S.\ \rpro\ pattern
than to the scaled S.S.\ \spro\ pattern, as noted previously
(e.g., \citealt{cowan02,roederer10a}).

A small \spro\ contamination, indicated by the combined
$r+s$ distribution in Figure~\ref{logepsplot}, 
might improve the fit to several elements in \hda.
In S.S.\ material a substantial portion of these elements 
(e.g., strontium, zirconium, cerium, lead;
$Z =$~38, 40, 58, 82) is attributed to the \spro,
so it is not surprising that they are the first
to show evidence of \spro\ enrichment.
There is no clear evidence of \spro\ contamination in \bd.
There is conflicting evidence of \spro\ contamination in \hdd;
the fit to several elements is improved by including
an \spro\ component (e.g., strontium, zirconium, cerium), while
for others (e.g., barium) it is not.
Furthermore, 
\hdd\ is a probable member of a stellar stream, and
abundances in the other members suggest that the star-forming regions of 
the stream's progenitor system were not polluted by significant
amounts of \spro\ material \citep{roederer10a}.
We advise against using these combined $r+s$ distributions
to quantify the relative amount of
$s$- and \rpro\ material present.
The predicted S.S.\ distributions are the cumulative yields of 
many nucleosynthesis events and may not be
representative of individual $s$- or \rpro\ events.
We conclude that \rpro\ nucleosynthesis is mainly
responsible for the heavy elements in these stars.

Despite the possibility of a small amount of \spro\ contamination,
several features are clear from Figure~\ref{logepsplot}.
The \spro\ contamination has a negligible effect on the
second and third \rpro\ peaks and
the heavy end of the rare earth domain.
The more pronounced odd-even effect observed in the
stellar distributions for strontium through cadmium, 
though mitigated by assuming an \spro\ contribution,
persists.
One platinum ($Z =$~78) non-detection in \hdd\ by \citet{roederer12}
indicates that platinum falls below the
scaled S.S.\ \rpro\ prediction, so the
third \rpro\ peak elements may be slightly deficient in this star.
Otherwise, these patterns are remarkably constant considering that the 
\rpro\ enrichment (e.g., [Eu/Fe]) spans 1~dex
in the three stars.

These results confirm that the tellurium abundances in halo stars
match the predicted S.S.\ \rpro\ residuals for the second \rpro\ peak.
They also indicate that tellurium is predominantly produced, along with
the rare earth elements, in the main component of the \rpro, as
has been predicted (e.g., \citealt{kratz07,farouqi10}).
The ratio of second-to-third \rpro\ peak abundances is an important 
benchmark for \rpro\ models, and our results confirm that the
\rpro\ operating in the early Galaxy produced ratios 
like those of the predicted S.S.\ \rpro\ residuals.
The S.S.\ and halo star abundances of elements lighter than tellurium 
have revealed the presence of material from 
additional nucleosynthesis mechanisms
(e.g., \citealt{wasserburg96,travaglio04,qian08})
and possibly
variations in the physical conditions of \rpro\ nucleosynthesis
(e.g., \citealt{truran02,roederer10c}).
Our results suggest that contributions 
from other nucleosynthesis mechanisms to the tellurium
in these stars are minor.

Tellurium is the heaviest element
whose \rpro\ production can be calculated 
largely based on experimental nuclear data \citep{dillmann03}.
Selenium, which has not been detected in \rpro\ 
enriched metal-poor stars, is another.
These are the first observations of
tellurium produced by the \rpro\ operating in the early Galaxy,
and they provide critical data for validating 
\rpro\ models with the least amount of nuclear physics uncertainty.

\acknowledgments

We appreciate the expert assistance of the STScI staff 
in obtaining these observations and the 
astronauts of STS-125 for their enthusiastic
return to \textit{HST} for Servicing Mission~4.
We also appreciate several helpful suggestions 
made by the referee, Charles Cowley.
Generous support for Program number 12268 was provided by NASA through
a grant from the Space Telescope Science Institute, which is operated by the
Association of Universities for Research in Astronomy, Incorporated, under
NASA contract NAS~5-26555.
I.U.R.\ is supported by the Carnegie Institution of Washington 
through the Carnegie Observatories Fellowship.
J.E.L.\ acknowledges support from NASA Grant NNX09AL13G.
T.C.B.\ and H.S.\ acknowledge partial support from grants PHY 02-16783 and
PHY 08-22648: Physics Frontier Center / Joint Institute for Nuclear
Astrophysics (JINA), awarded by the U.S.\ National Science Foundation.
H.S.\ acknowledges additional support from NSF Grant PHY-1102511.
C.S.\ acknowledges support from NSF Grant AST~09-08978.

{\it Facilities:} 
\facility{HST (STIS)}

\end{document}